\documentclass[Circuits, Systems, and Signal Processing, final]{svjour3}
\usepackage{color}
\usepackage{graphicx}
\usepackage{mathptmx}
\usepackage{amsmath}
\usepackage{amsfonts}
\usepackage{epstopdf}
\usepackage{amssymb}
\usepackage{algorithm}
\usepackage{algorithmicx}
\usepackage{algpseudocode}
\algnewcommand\algorithmicoutput{\textbf{Output:}} 
\algnewcommand\Output{\item[\algorithmicoutput]}
\algnewcommand\algorithmicinput{\textbf{Input:}} 
\algnewcommand\Input{\item[\algorithmicinput]}

 \journalname{Circuits, Systems, and Signal Processing, \ \ Springer -- Birkhäuser, \ }

\begin{document}

\title{RANSAC-Based Signal Denoising  \\ Using
	Compressive Sensing}

\author{Ljubi\v{s}a Stankovi\'{c}, 
	 Milo\v{s} Brajovi\'{c}, \\
 Isidora Stankovi\'{c},
Jonatan Lerga, \\
Milo\v{s} Dakovi\'{c}
}

\institute {L. Stankovic, M. Brajovic, I. Stankovic, and M. Dakovic are with the University of Montenegro, J. Lerga is with the Faculty of Engineering  and Center for Artificial Intelligence and Cybersecurity, University of Rijeka, Croatia. }
	  

\maketitle

\begin{abstract}
In this paper, we present an approach to the reconstruction of signals exhibiting sparsity in a transformation domain, having  some heavily disturbed samples. This sparsity-driven signal recovery exploits a carefully suited random sampling consensus (RANSAC) methodology for the selection of an inlier subset of samples. To this aim, two fundamental properties are used: a signal sample represents a linear combination of the sparse coefficients, whereas the  disturbance degrade original  signal sparsity. The properly selected samples are further used as measurements in the sparse signal reconstruction, performed using algorithms from the compressive sensing framework. Besides the fact that the disturbance degrades signal sparsity in the transformation domain, no other disturbance-related assumptions are made -- there are no special requirements regarding its statistical behavior or the range of its values. As a case study, the discrete Fourier transform (DFT) is considered as a domain of signal sparsity, owing to its significance in signal processing theory and applications. Numerical results strongly support the presented theory. In addition, exact relation for the signal-to-noise ratio (SNR) of the reconstructed signal is also presented. This simple result, which conveniently characterizes the RANSAC-based reconstruction performance, is numerically confirmed by a set of statistical examples.


\end{abstract}

\section{Introduction}

In recent years, the reconstruction of sparse signals, based on a reduced set of random measurements, attracted significant research interest \cite{donoho2006,candes2006,flandrin,cosamp,LJSTut,LJSTdems,impulsiveCS,ImpulsiveHT,impulsiveAudio,Zhao,Rui,AccessCSquant,nonuniformCS_geo,D1,Imp,Machinparsuit1,djurovic2014quasi,MJF,20,LJTSP,16a,CS,LJSAut,LJSaudio,Isi,Orovic,Toroko,Khan}. Within the compressive sensing (CS) framework, a rigorous mathematical foundation has been established to support this type of reconstruction, including the conditions that  guarantee a successful and unique reconstruction result  \cite{candes2006,LJSTut}. The most important requirement imposed by this theory is that the signal must be sparse in a particular transformation domain. Such signals are characterized by a small number of non-zero elements in the sparsity domain as compared to the full signal length, and do appear in various applications \cite{LJSTut}.

Signal samples in the time-domain, being linear combinations of the sparsity domain coefficients, can be considered as the measurements within the CS framework. The number of samples required for the reconstruction is closely related to the number of nonzero coefficients in the sparse domain
\cite{donoho2006,candes2006,LJSTut}. It has been repeatedly confirmed that in the case of heavily disturbed samples, it is preferable to omit these samples (outliers) in both  signal analysis and  processing \cite{16a,7a}.
 In the CS and sparse signal processing context, this means that the uncorrupted or low-corrupted samples (inliers) are considered as available measurements, whereas the heavily corrupted samples are considered as unavailable, and are reconstructed by applying some of the CS-based recovery algorithms. This further implies that  a detection method shall be applied to identify the highly corrupted samples prior to their reconstruction \cite{LJSTut}. One such  methodology, based on the signal sparsity principles, is presented in this paper.
In contrast to the methods that combine the robust estimation and the CS
based signal recovery, the proposed method can provide an exact reconstruction if the reduced set of inliers is noise-free, rather than to obtain a filtered approximation of the original signal   \cite{djurovic2014quasi,ramadan2012efficient,RoenkoLukin,SVV,dadouchi2013automated,ahmad2012robust,you2013novel}.

The proposed algorithm  is based on a random selection of subsets of signal samples (random sampling consensus - RANSAC) and a detection of the event when a disturbance-free (or low-corrupted) subset is selected. The RANSAC-based approach improves our detection and reconstruction based on sparsity measures, reported in \cite{impulsiveCS}, in the way that it provides a criterion to properly choose the best and largest possible set of adequate measurements -- inliers, which are suitable for the CS-based reconstruction of corrupted samples -- outliers. The SNR analysis in the CS-based reconstruction  \cite{LJSTut,LJSAut,LJSTerr} shows that  the quality of the reconstruction is directly influenced by the number of available samples. The presented RANSAC-based methodology is  additionally equipped with the analysis of noise influence to  the reconstructed signal, using an exact and simple relation for the SNR in the resulting signal.

The paper is organized as follows. After the introduction, the background of the compressive sensing theory and the RANSAC overview are  presented in Section \ref{basic}. The denoising procedure based on the RANSAC is presented in  Section \ref{detectionalg}, which also includes the discussion regarding the numerical complexity. The presented theory is verified numerically in Section \ref{numericals}, while the paper ends with some concluding remarks.

\section{Definitions}
\label{basic}

Consider a signal $x(n)$ with $N$ samples in the discrete-time domain. Assume
that the sparsity domain of the signal is the discrete Fourier transform (DFT)
domain. The signal and the DFT coefficients are related via%
\begin{align}
X(k)=\mathrm{DFT}\{x(n)\}  &  =\sum_{n=0}^{N-1}x(n)\varphi_{k}^{\ast}(n),\label{defTR}\\
x(n)=\mathrm{IDFT}\{X(k)\}   &  =\frac{1}{N}\sum_{k=0}^{N-1}X(k)\varphi_{k}(n) \label{defTRi}%
\end{align}
or $\mathbf{X}=\mathbf{Wx}$ and $\mathbf{x}=\mathbf{W}^{-1}\mathbf{X}$, where the basis functions take the form $\varphi_{k}(n)=\exp(j2\pi nk/N)$. In  the matrix notation,
the $N$-dimensional vector $\mathbf{X}$ has elements $X(k)$, vector
$\mathbf{x}$ has elements $x(n)$, and the $N\times N$ matrix $\mathbf{W}$ is
with elements $\varphi_{k}^{\ast}(n)$, $k=0,1,\dots,N-1$, $n=0,1,\dots,N-1$. 

The time-domain representation of a signal which
is sparse in the DFT domain reads
\begin{equation}
x(n)=\sum\limits_{i=1}^{K}A_{i}\varphi_{k(i)}(n) , \label{signal_def}%
\end{equation}
where $A_i$, $i=1,2\dots,K$, represent the nonzero elements at the transformation  domain positions $k(i)$,  $i=1,2,\dots,K$. The sparsity of this signal is $K$, $(K\ll N)$.

Assume that there are $M$ available (inlier) samples at the instants
$n_{i}$, $i=1,2,$\dots$,$ $M$. In the CS theory, the linear combinations of the sparsity coefficients are referred to as the measurements. Therefore, the available signal samples can be considered as the CS measurements, $y(i)=x(n_{i})$. The measurement vector
 is denoted by
$\mathbf{y}$. Its elements are
\[
y(i)=x(n_{i})=\frac{1}{N}\sum_{k=0}^{N-1}X(k)\varphi_{k}(n_{i}) \label{availsig},~~i=1,2,\dots,M.
\]
The matrix notation for this system of $M$ linear
equations is 
\begin{equation}
\mathbf{y=AX}. \label{masuremetsyst}
\end{equation} 
The measurement matrix $\mathbf{A}$ is obtained from the inverse DFT matrix
$\mathbf{W}^{-1}$, by eliminating the rows corresponding to the
unavailable/outlier samples.

The goal of the CS approach is to reconstruct the complete signal $x(n)$, $n=0,1,2,\dots,N-1$, from the reduced (compressively sensed) subset of (inlier) samples at the instants $n_{i}$, $i=1,2,\dots,M$, with $M<N$. 
 This problem cannot be solved in a direct way, since there are $M<N$ equations for $N$ unknown transformation domain elements of $X(k)$. When the signal sparsity is assumed, the sparsity measure minimization is added.  Among an infinite number of solutions for the underdetermined system (\ref{masuremetsyst}),  $\mathbf{y=AX}$, we look for the one with the sparsest possible representation in the transformation domain while satisfying the measurement equations. The constrained minimization problem formulation is, 
\begin{equation}
\min \left\Vert \mathbf{X}\right\Vert _{0}\text{ subject to }\mathbf{y=AX}.
\label{DEF_L0}%
\end{equation}
The solution to this problem exists (if some conditions are met) and it produces the signal transformation elements $X(k)$ and the signal samples at all instants. 
 
Simple counting of the nonzero values of $X(k)$ is achieved using the so-called
$\ell_{0}$-norm $\left\Vert\mathbf{X}\right\Vert _{0}$. However, the solution based on the $\ell_{0}$-norm is an NP-hard optimization problem. Its
calculation complexity is of order $\binom{N}{K}$. In theory, the NP-hard
problems can be solved by an exhaustive search. However, as the problem
parameters $N$ and $K$ increase, the computational time increases and the problem
becomes unsolvable in a realistic time frame. 

Many methods are recently developed for the CS-based reconstruction, by solving the minimization problem (\ref{DEF_L0}) or its various equivalent forms (a review of some of them can be found in \cite{LJSTut}). In this paper,
will use a very simple and efficient algorithm that belongs to the class of matching pursuit algorithms. This algorithm will be described in Section \ref{RansacCS}. 

\subsection{RANSAC Review}

The random sample consensus (RANSAC) is used for linear regression when the outliers in the data are expected. Consider a set of data $x(n_i)$ sampled at random instants $n_i$. Assume that the true data values fit a linear model, $x=an+b$. Since a large number of  outliers is expected, most of the data samples can be far from the linear model. In the RANSAC approach, we will:
\begin{enumerate}
	\item
	Assume a small subset $\mathbb{S}$ with $S$ randomly selected samples $x(n_i)$ at $n_i \in \mathbb{S}$. 
	\item \label{point_2} The samples with indices in $\mathbb{S}$ are used to estimate the linear regression model parameters,
	\begin{gather*}
	a \! \! \sum_{n_i\in \mathbb{S}}n^2_i+b \sum_{n_i\in \mathbb{S}}n_i=\!\!\sum_{n_i\in \mathbb{S}}n_ix(n_i) \text{  \ \ and\ \ \ \ \  }
	a \! \! \sum_{n_i\in \mathbb{S}}n_i+b N=\!\!\sum_{n_i\in \mathbb{S}}x(n_i).
	\end{gather*} 
	\item   
	After the parameters $a$ and $b$ are determined,
	the line $$x=an+b$$ is plotted.  
	The distances $d_n$ of all data points $(n,x(n))$, $n=0,1,\dots,N-1$, from this line are calculated, $d_n=|an+b-x(n)|/\sqrt{1+a^2}$.
	\item 
	If a sufficient number of data points is such that their distance from the model line is lower than an assumed distance threshold $d$, then all these points are included into a new set of data $$\mathbb{D}=\{(n_i,x(n_i))| \ \ \ d_{n_i}\le d \},$$
	and the final parameters $a$ and $b$ (for machine learning or prediction) are calculated with all data from $\mathbb{D}$. 
	\item 
	If there was no sufficient number of data points within the distance $d$, a new random small set  of data, $\mathbb{S}$, is taken and the procedure is repeated from point \ref{point_2}.
	\item The procedure ends when the desired number of data points within $\mathbb{D}$ is achieved or the maximum number of trials is reached.
\end{enumerate}  

\section{Reconstruction of Sparse Signal with Outliers}
\label{detectionalg}
Consider a signal $x(n)$, $0\leq n\leq N-1$, that is sparse in a transformation
domain with a sparsity $K\ll N$. Assume that the signal is noisy with a small noise $\varepsilon(n)$ (causing inliers) in all samples. Assume that $I$ samples of the signal $x(n)$,
at unknown positions $n\in\mathbb{N}_{I}$, are corrupted with an impulsive disturbance
$\nu(n)$. The disturbing signal $\nu(n)$ can be modeled, for example, as
$\nu(n)=0$ for $n\notin\mathbb{N}_{I}$ and $\nu(n)$ assumes
arbitrary high values for $n\in\mathbb{N}_{I}$. The original signal can be 
recovered if a sufficient number of the inlier samples exists.

The sufficient number of inliers is directly related to the full
recovery conditions studied in the CS theory
\cite{donoho2006,candes2006,cosamp}. A rough estimation of the smallest number of samples that should be used in the reconstruction of $K$ sparse signal can be
made based on the statistical results presented in \cite{impulsiveCS}.
The exact reconstruction conditions are commonly defined by the restricted isometry property (RIP), using the spark, or the coherence  of the measurement matrix \cite{donoho2006,candes2006,LJSTut,LJSTdems}. However, these 
	conditions are either computationally unfeasible or too pessimistic. In numerical tests, we followed   the set of practical guidelines from \cite{donoho2006,LJSTdems} for situations when one could expect perfect recovery from a partial Fourier matrix using convex optimization. It is suggested that in the case when  $K \le M/5$, the recovery rate is highly reliable.  It is worth mentioning that computationally feasible results for the reconstruction uniqueness of signals sparse in the DFT domain  are presented in \cite{Uniq}, \cite{LJSBOOK}.
	
	 It is important to emphasize that the use of lower values for $M$, when some unsuccessful reconstructions are expected to happen, will not be problematic for the RANSAC, since this methodology, by definition, searches for successful reconstruction, as it will be shown in the next section.

\subsection{Concentration Measure-Based  Denoising}

To solve the stated problem,   in our previous work \cite{impulsiveCS,ImpulsiveHT,impulsiveAudio,ISTdems,LJSTerr}, we exploited the idea of
eliminating random subsets of samples and performing the reconstruction using
the remaining samples. For each realization, the sparsity measure of the
recovered signal is used for the detection of the full recovery event. By using a
sparsity measure close to the $l_{0}$-norm, all realizations containing disturbed
samples will produce a sparsity measure close to the total number of
samples $N$. This is expected, since a disturbance at any signal sample, will affect all coefficients in the sparsity domain \cite{impulsiveCS}. As an illustration, consider a disturbance at $n_i$, of the form $\nu(n)=\nu_0\delta(n-n_i)$, which is added to the signal sample $x(n)$ at $n_i$. Its DFT, $ X_{\nu}(k)=\nu_0\exp(j2\pi n_i k/N)$, $k=0,1,\dots, N-1$, is spread over all DFT coefficients of the original signal. In many CS approaches, $\ell_1$-norm is used as a measure of signal sparsity \cite{Ljubisa}. Noise influence on sparsity measures has also been investigated in \cite{Jokanovic,viktor_impulsive}.

In the case when only the uncorrupted samples are used in the
reconstruction, the sparsity measure is of order $K$. It is much lower than the
total number of samples $N$. Therefore, by setting a threshold $T_{\mu}$ within
$K<T_{\mu}<N$, we can detect a full recovery event.

 The main idea behind this approach can be significantly improved, with respect to the detection criterion and the final signal estimation, using  an increased number of samples provided by the RANSAC method, a classical method which has been extensively applied in machine learning \cite{ML1,ML2,ML3} and  signal processing as well \cite{igorRANSAC}.

\subsection{RANSAC-based CS Signal Denoising}\label{RansacCS}
In this section, we present the RANSAC-based CS denoising approach. The considered  signal model is
\begin{gather}
y(n)=x(n)+\varepsilon(n)+\nu(n)
\end{gather}
where $x(n)$ is a signal sparse in the DFT domain, $\varepsilon(n)$ is Gaussian noise with variance $\sigma^2_{\varepsilon}$, and $\nu$ is the impulsive noise. The Gaussian noise is considered as a small noise and the samples with this kind of noise as inliers.  The impulsive noise causes outliers in the signal samples.

For the RANSAC-based denoising, we will use the following algorithm:
\begin{enumerate}
	\item
	The considered signal is $K$-sparse. Select a small subset $\mathbb{S}$ with $S$ randomly selected samples $x(n)$ at $n \in \mathbb{S}$, such that the reconstruction for $K$-sparse signal is possible. 
	\item The samples with indices in $\mathbb{S}$ are used to reconstruct the signal, $x_R(n)$, at all instants $n=0,1,\dots,N-1$.
	An approach to solve the CS reconstruction problem is  a two-step strategy as follows \cite{LJSTut}: 
	\begin{itemize}
		\item[] Step 1: Detect the positions of nonzero elements in the sparsity domain,
		\item[] Step 2: Apply an algorithm for reconstruction with known positions of nonzero elements (use the positions from Step 1).
	\end{itemize}	
	
\smallskip
	The measurements, $x(n)$, are characterized by a linear nature: they are obtained as linear combinations of the sparsity domain elements, $X(k)$, with the corresponding rows of the measurement matrix, $\mathbf{A}$, acting as weights. This further implies that a back-projection of the measurements, $\mathbf{y}$, to the measurement matrix, $\mathbf{A}$, defined by
	\begin{equation}
	\mathbf{X}_0=\mathbf{A}^H \mathbf{y}=\mathbf{A}^H \mathbf{A} \mathbf{X}\label{Initi}
	\end{equation}
	can be used to estimate the positions of nonzero elements in $\mathbf{X}$.  
	
	In an ideal case, the matrix $\mathbf{A}^H \mathbf{A}$ should ensure that the initial estimate, $\mathbf{X}_0$, contains exactly $K$ elements at positions $\{k_1,k_2,\dots,k_K\}$, for which the magnitudes are larger than the largest magnitude at the remaining positions. Under such condition, by taking the positions of these largest magnitude elements in $\mathbf{X}_0$ as the set $\{k_1,k_2,\dots,k_K\}$, the algorithm for the known nonzero element positions, can be applied to reconstruct the signal using the pseudo-inversion as
	\begin{equation}\mathbf{X}_K=(\mathbf{A}^H_{K}\mathbf{A}_{K})^{-1}\mathbf{A}_{K}^H\mathbf{y}=\textrm{pinv}(\mathbf{A}_{K})\mathbf{y}, \label{pinvSol}
	\end{equation} 
	where $\mathbf{A}_{K}$ is the matrix obtained from the measurement matrix $\mathbf{A}$, by keeping the columns which correspond to the indices $\{k_1,k_2,\dots,k_K\}$.
	
	The reconstructed signal is 
	$$x_R(n)=\mathrm{IDFT}\{X_{K0}(k)\}$$
	where $X_{K0}(k)$ are zero-valued at all $k$ except $k\in \{k_1,k_2,\dots,k_K\}$, where $X_{K0}(k_i)=X_K(i)$.
	
	 The procedure can be iteratively implemented \cite{LJSTut}. In the iterative procedure, the largest signal component is detected and estimated first (as it were $K=1$). The reconstructed component is subtracted from the available measurements, and the modified measurements are used to detect the position of the second-largest component. The two largest components are re-estimated together and subtracted from the available measurements. The procedure is continued until the subtraction of the estimated components from the available measurements does not produce sufficiently small (stopping criterion) value or until an assumed sparsity, $K$, is reached. 
	 
	 The iterative implementation is particularly suitable in the  case when the nonzero coefficient values significantly  differ. By reconstructing the detected large components, during the iterations, and removing their influence on the previously non-detected  components, smaller components will emerge \cite{LJSTut}.

	\item   
	After the signal is reconstructed, the distance $d_n$ of all signal samples $(n,x(n))$, $n=0,1,\dots,N-1$, from the estimated signal $x_R(n)$ are calculated, $d_n=|x_R(n)-x(n)|$.
	\item 
	If a sufficient number of signal values is such that their distance from the reconstructed signal (model) is lower than an assumed threshold, for example, for the Gaussian distributed inliers $d=2.5\sigma_{\varepsilon}$, then all these points are included in the new set of signal values $$\mathbb{D}=\{(n,x(n)) \ | \ \ \ d_n\le d \},$$
	and the final reconstruction is calculated \textit{with all data from $\mathbb{D}$}. Note that the robust estimation of the standard deviation can be done using  the median absolute deviation (MAD), defined by
	$$MAD_x={\underset{m=0,1,\dots,N-1}{\mathrm{median}}\large\{\big|x(m)-\underset{n=0,1,\dots,N-1}{\mathrm{median}}\{x(n)\}\big|\large\}}.$$
	The MAD value is related to the sample standard deviation as $MAD_x=0.6745\sigma_x$ (for the Gaussian random variable). The real and imaginary parts are considered separately, for complex valued signals. 
	\item 
	If there was no sufficient number of signal values within the distance $d$, that is, $\mathrm{card}\{\mathbb{D}\}<T=3N/4$, a new random small set  of signal samples, $\mathbb{S}$, is taken and the procedure is repeated from Step 2. The value $T$ is assumed based on the expected number of outliers. The assuming $T=3N/4$ means that we do not expect more than $N/4$ outliers. 
	
	\item The procedure is stopped when the desired number of data points within $\mathbb{D}$ is achieved, $\mathrm{card}\{\mathbb{D}\}\ge T=3N/4$ or the maximum number of trials $N_{max}$ is reached.
\end{enumerate}  

The previously described denoising approach is summarized in Algorithm 1. It exploits a matching pursuit CS reconstruction procedure, \Call{CSrec}{$\cdot$}, summarized in Algorithm 2. 

 The presented RANSAC-based denoising algorithm will produce the same results if other CS reconstruction methodologies are used, such as, for instance, the Bayesian CS reconstruction or the iterative hard thresholding (IHT). An overview of these procedures, along with their algorithmic presentation, suitable for implementation, can be found in \cite{LJSTut,AccessCSquant}.

\begin{algorithm}[!h] \label{ReconMetp}
	\caption{RANSAC CS Denoising Algorithm}
	\begin{algorithmic}[1]
		\Input Noisy signal $\mathbf{x}$, RANSAC set size $M$, bound for inliers $d$, threshold for the consensus number of samples $T$, maximum number of iterations $N_{max}$, signal sparsity $K$
		
		\Statex
		\State $D \gets 0$, $N_{it} \gets 0$
		\While {$D<T$ and $N_{it}\le N_{max}$}
		\State $N_{it} \gets N_{it}+1$
		\State $\mathbb{S} \gets \mathrm{randperm}(N,M)$, \Comment $M$ random numbers from the first $N$ natural numbers
		\State $\mathbf{A} \gets $ rows of the  inverse DFT matrix $\mathbf{W}^{-1}$ selected by 
		the set $\mathbb{S} $ 
		\State $\mathbf{ {y}}\gets$ elements of $\mathbf{x}$ selected by the set $\mathbb{S}$
		\State  $\mathbf{X} \gets$ \Call{CSrec} {$\mathbf{y},\mathbf{A},K$} 
		\State $\mathbf{x}_R=\mathbf{W}^{-1}\mathbf{X}$
		\State $\mathbb{D}=\mathrm{find}(|\mathbf{x}-\mathbf{x_R}|<d)$, \State $D=\mathrm{card}(\mathbb{D})$, \Comment the number of  elements in $\mathbb{D}$
		\EndWhile
		\Statex 
		\Statex  Repeat the CS reconstruction with the consensus set $\mathbb{D}$:
		\Statex 
		\State $\mathbf{A} \gets $ rows of the  inverse DFT matrix $\mathbf{W}^{-1}$ selected by 
		set $\mathbb{D} $ 
		\State $\mathbf{ {y}}\gets$ elements of $\mathbf{x}$ selected by the set $\mathbb{D}$
		
		\State  $\mathbf{X} \gets$ \Call{CSrec} {$\mathbf{y},\mathbf{A},K$} 	
		\State $\mathbf{x}_R\gets \mathbf{W}^{-1}\mathbf{X}$
		\Output
		Reconstructed denoised signal $\mathbf{x}_R$ 
	\end{algorithmic}
\end{algorithm}

\begin{algorithm}[!h] \label{CSrec}
	\caption{Matching Pursuit CS Reconstruction}
	
	\begin{algorithmic}[1]
		\Statex
		\Function{CSrec}{$\mathbf{ {y}}, \mathbf{A},K$}
		
		%
		\State $\mathbb{K} \gets \emptyset$, \ \ $\mathbf{e} \gets \mathbf{y} $, 
		\For {$i=1$ to $K$} 
		\State $k \gets $ position of 
		the largest 
		value in $|\mathbf{A}^H\mathbf{e}|$
		\smallskip
		\State $\mathbb{K} \gets \mathbb{K} \cup k $
		\State $\mathbf{A}_K \gets $ columns of the  measurement  matrix $\mathbf{A}$ selected by the 
		set $\mathbb{K} $
		\State $\mathbf{X}_K \gets \operatorname{pinv}(\mathbf{A}_K)\mathbf{y}$
		\State $\mathbf{y}_K \gets \mathbf{A}_K\mathbf{X}_K$ 
		\State $\mathbf{e} \gets \mathbf{y} -  \mathbf{y}_K$
		\EndFor
		\State $\mathbf{X} \gets \mathbf{0}$, \ \ $\mathbf{X} \gets \mathbf{X}_K$ for $k \in  \mathbb{K}$
		
		\State \Return  $\mathbf{X}$ 

		\EndFunction
	\end{algorithmic}
\end{algorithm}

\subsection{Calculation Complexity}
 The main difference between RANSAC and the standard CS methods is in the iterative procedure with random subsets of samples. Since this  is the main factor in
the computational complexity of this method, we will find the probability that
within $M$ randomly selected observation samples, there are no outliers.

The probability that the first randomly chosen sample is not affected by an outlier is $(N-I)/N$ since there are $N$ samples in total and $N-I$ of
them are inliers. We continue the process of random sample selection, and the probability that both the first and the second chosen samples are not outliers is $\frac{N-I}{N}\frac{N-I-1}{N-1}$. In this way, we can
calculate the probability that all of $M$ randomly chosen samples $x(n)$ at the 
positions $n\in\mathbb{S}$ are not outliers. This
probability is%

\begin{equation}
P(M,N,I)=\prod_{i=0}^{M-1}\frac{N-I-i}{N-i}. \label{vjerovatnoca}%
\end{equation}
Since $\frac{N-I-i}{N-i}<1$, we can see that the probability $P(M,N,I)$ decreases
as the number of terms in the product increases. Thus, in this kind of
reconstruction it is important to keep the number of samples $M$ in the
observation set as low as possible, while satisfying the CS reconstruction conditions. 

In general, for an expected number of pulses $I,$ the expected number of
random realizations to achieve at least one outlier-free reconstruction
using a subset of $M$ samples is 
$$N_{it}=\frac{1}{P(M,N,I)}.$$

In classic literature dealing with the RANSAC, it is common to use the following calculation for the expected number of the iterations to get an outlier-free realization. The probability that one randomly selected sample is inlier is $(N-I)/N$. It is then assumed that this probability can be used for $M$ samples. The probability that there is at least one outlier  in $M$ samples is $[1-((N-I)/N))^M]$. Finally, the probability of an outlier-free realization in $N_{it}$ such trials is      
$$P=1-[1-((N-I)/N))^M ]^{N_{it}}$$
where
$$N_{it}=\frac{\ln(1-P)}{\ln(1-((N-I)/N))^M)},$$  with the given probability $P$.
This calculation is correct if the approximation 
$$\frac{N-I-M}{N-M} \approx \frac{N-I}{N},$$ holds, otherwise instead of $((N-I)/N))^M$ we should use $P(M,N,I)=\prod_{i=0}^{M-1}(N-I-i)/(N-i)$ to get the correct result.

\subsection{Expected Signal-to-Noise Ratio}\label{SNRD}

Assume that the RANSAC has produced the correct result, that is, the reconstruction which is not influenced by the outliers.  This means that the final reconstruction is done using all consensus samples in the set $\mathbb{D}$. This set of samples contains only inliers (with the Gaussian noise). Then, we can use a simple formula for the output SNR in the $K$-sparse signal, reconstructed from $D$ samples,
 $SNR_{out}$, derived in \cite{LJSTut,LJSAut,LJSTerr}
\begin{equation}
SNR_{out}=SNR_{in0}+10\log\bigg(\frac{D}{K}\bigg).
\end{equation} 
Since the outliers are eliminated before the reconstruction, the input SNR is denoted by $SNR_{in0}$ and it includes the noise in inliers only. For reference, we will provide the total input SNR in notation $SNR_{in}$, which includes outliers as well. We will also provide the value of the SNR in the  mid-result, when the sample consensus is reached in a small subset $\mathbb{S}$ with $M$ samples. This SNR is denoted by $SNR_{out0}$ and its relation to the input SNR is 
\begin{equation}
SNR_{out0}=SNR_{in0}+10\log\bigg(\frac{M}{K}\bigg).
\end{equation} 
The improvement achieved by the RANSAC-based approach with respect to the random selection of the subsets and the concentration measure-based selection of the reconstruction set, is equal to 
\begin{equation}
SNR_{out}-SNR_{out0}=10\log\bigg(\frac{D}{K}\bigg)-10\log\bigg(\frac{M}{K}\bigg)=10\log\bigg(\frac{D}{M}\bigg).
\end{equation}    
This improvement can be significant, having in mind that we have to keep $M$ as small as possible, in order to reduce the expected number of iterations $N_{it}$, which is crucial for the calculation complexity, for both the measure-based denoising and the RANSAC based denoising. The number of consensus samples, $D$, can be as high as the number of inliers, meaning that the ratio $D/M$ could be very large. 

\section{Numerical and Statistical Examples}
\label{numericals}
The results from the last two sections will now be illustrated on  numerical and statistical examples with  sparse noisy signals, containing inliers and outliers, reconstructed using the CS version of the RANSAC.

\smallskip
\noindent\textbf{Example 1:}  
A general form of a noisy signal, sparse in the DFT domain, is considered
 \begin{gather}
y(n)=x(n)+\varepsilon(n)+\nu(n),
\end{gather}
where $$x(n)=\sum_{i=1}^{K}A_ke^{j(2\pi k_i/N+\phi_1)}$$
is a signal sparse in the DFT domain with  $K$ randomly positioned nonzero amplitudes and random phases. The Gaussian complex-valued noise $\varepsilon(n)$ is with $\sigma_{\varepsilon}=0.5$. The impulsive Cauchy noise $\nu(n)$ is formed as $\nu=3\varepsilon_1(n)/\varepsilon_2(n)+j3\varepsilon_3(n)/\varepsilon_4(n)$, where $\varepsilon_i(n),\,i=1,2,3,4$ are unit variance, zero-mean, Gaussian noises. The impulsive noise is added in $I$ signal samples. In the denoising of this signal, the CS form of the RANSAC is applied with random subsets of $M$ samples, where $M$ is small enough to keep the lowest possible calculation complexity, but sufficient to provide the correct reconstruction of a $K$ sparse signal with acceptable probability.   

Several cases, for specific values of the total number of samples, $N$, sparsity, $K$, number of outliers, $I$, and the size of the RANSAC subset, $M$, are considered.

\begin{figure}[htbp]
	\begin{center}
		\includegraphics[
		scale=1
		]{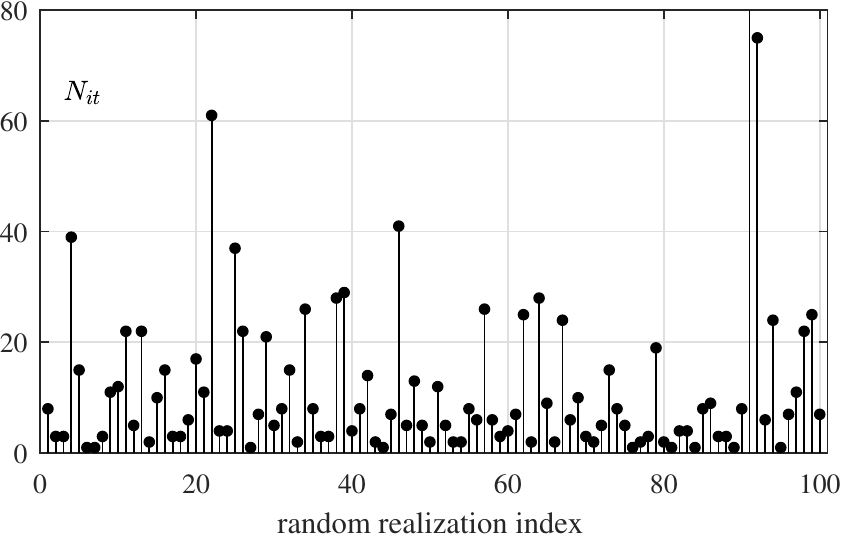}
	\end{center}
	\caption{The number of iterations $N_{it}$ in the reconstruction of a signal using the RANSAC, with $I=16$ out of $N=128$ samples being
		affected by an impulsive Cauchy disturbance. The average value is $11.99$.}%
	\label{RANSAC_CS_Nit}%
\end{figure}

\begin{figure}[htbp]
	\begin{center}
		\includegraphics[
		scale=1
		]{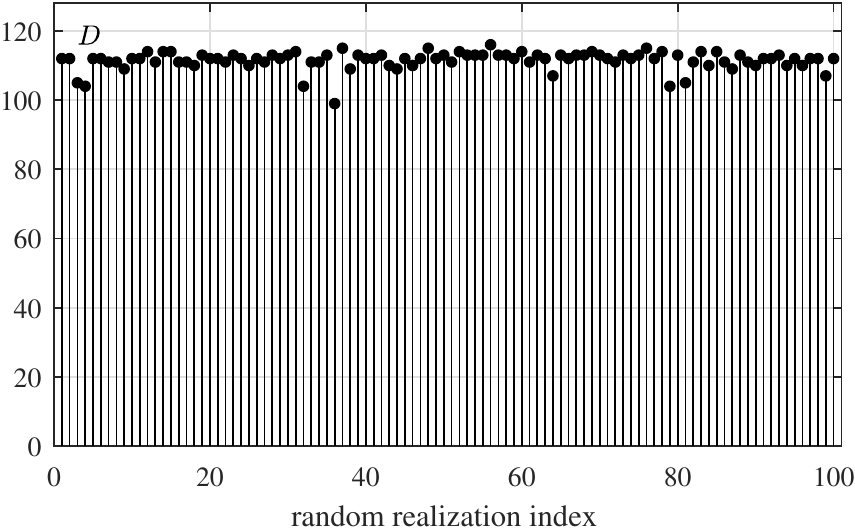}
	\end{center}
	\caption{The number of inliers in the solution in the reconstruction of a signal  using the RANSAC, with $I=16$ out of $N=128$ samples being
		affected by an impulsive Cauchy disturbance. The average value is $111.33$. }%
	\label{RANSAC_CS_DN}%
\end{figure}

\begin{figure}[htbp]
	\begin{center}
		\includegraphics[
		scale=1
		]{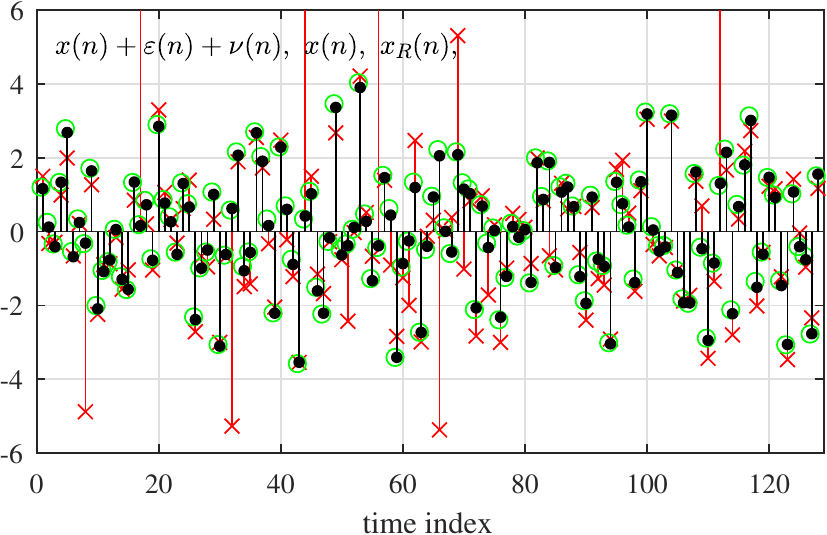}
	\end{center}
	\caption{Reconstruction of a signal with $I=16$ out of $N=128$ samples  using the RANSAC, being
		affected by an impulsive Cauchy disturbance. One random realization is shown for the illustration. The noisy signal is marked by the red crosses, the reconstructed signals denoted by the green circles, while the original signal is shown with the black dots. }%
	\label{RANSAC_CS}%
\end{figure}

\begin{figure}[htbp]
	\begin{center}
		\includegraphics[
		scale=1
		]{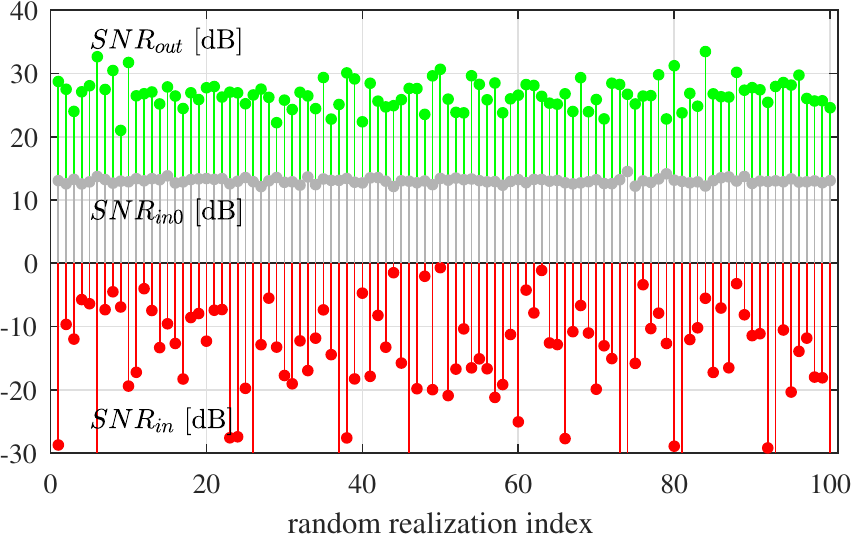}
	\end{center}
	\caption{The signal-to-noise ratio values in the reconstruction of a sparse signal  using the RANSAC with $I=16$ samples (out of $N=128$) being
		affected by an impulsive Cauchy disturbance. The average values are $SNR_{in}=-16.73$ dB for the input signal-to-noise, $SNR_{in0}=13.00$  dB for the input signal to Gaussian noise (when the impulsive noise is not counted), $SNR_{out}=26.46$ dB for the RANSAC reconstructed (output) signal. }%
	\label{RANSAC_CS_SNRoutD}%
\end{figure}

The case for $N=128$, $K=5$, $I=16$, and $M=32$ is presented in detail with illustrations in Fig. \ref{RANSAC_CS_Nit}, Fig. \ref{RANSAC_CS_DN},  Fig. \ref{RANSAC_CS}, and Fig. \ref{RANSAC_CS_SNRoutD}.
This experiment is performed $100$ times with different random realizations of signals, $x(n)$, and noises, $\varepsilon(n)$ and $\nu(n)$. The values of the number of iterations, $N_{it}$, in every realization is calculated and shown in Fig. \ref{RANSAC_CS_Nit}. The final number of signal samples in the sample consensus, $D$, that is used in the final signal reconstruction is given in  Fig. \ref{RANSAC_CS_DN}. One of the realizations of the noisy signal, original signal, and the reconstructed signal is shown in Fig. \ref{RANSAC_CS}. The noisy signal is marked by the red crosses, the reconstructed signal is denoted by the green circles, while the original signal is shown with the black dots. The SNR for the input signal $x(n)+\varepsilon(n)+\nu(n)$, $SNR_{in}$, for the input signal without impulsive noise $x(n)+\varepsilon(n)$, $SNR_{in0}$, and for the reconstructed signal $x_R(n),$  $SNR_{out}$, for every realization, is given in  Fig. \ref{RANSAC_CS_SNRoutD}.

The statistical results averaged over all considered realizations are given in Table \ref{tableI}.

\begin{table}[htbp]
	\centering
		\begin{tabular}{|l|r|r|r|r|r|r|}
			\hline 
			& $N_{it}$ & $SNR_{in} $ & $SNR_{in0}$ & $SNR_{out0}$ & $SNR_{out}$ &  $D$ \\ \hline
			
			$I=8$ & 2.94 &  -8.81 &  13.01 &  16.23 &  27.07 & 119.46 \\ \hline
			
			$I=16$ & 11.99 & -16.73 &   13.00 &  14.39 &   26.46 &  111.33 \\ \hline
			
			$I=24$  & 57.66 & -17.53 &  13.01 &   13.29 &   25.91 &  103.56 \\ \hline
		\end{tabular}
	\smallskip
	\caption{Results for the case with $N=128$, $M=32$, $K=5$.}
\label{tableI}
\end{table}

The theoretically expected improvement (Section \ref{SNRD}) in the SNR (omitting the impulsive noise in the input signal) for the  case $I=8$, in Table \ref{tableI}, is 
$$SNR_{out}-SNR_{in0}=10\log\bigg(\frac{D}{K}\bigg)=10\log\bigg(\frac{119.46}{5}\bigg)=13.77 \text{ dB}$$
For other two cases, $I=16$ and $I=24$, we get $$SNR_{out}-SNR_{in0}=13.46 \text{ dB}$$
 and $SNR_{out}-SNR_{in0}=13.14$ dB, respectively. This is in high agreement with the statistical data in Table \ref{tableI}.

\smallskip
\noindent\textbf{Example 2:}  
Since the impulsive Cauchy noise may take some small values as well (some of the assumed $I$ outliers may happen to be inliers, in reality), the expected number of iterations is smaller than the theoretically obtained result given by (\ref{vjerovatnoca}). In order to check the expected number of iterations, $N_{it}$, against its theoretical value, avoiding the ambiguity of possible Cauchy noise inliers, we will provide that all $I$ signal samples are certainly outliers. The impulsive noise is modified as $\nu(n) \to \nu(n)+100$, to be sure that all of these samples are the outliers. Then, we have repeated the same experiment with $100$ realizations and obtained $N_{it}=10.74$, while the theory in (\ref{vjerovatnoca}) predicts $P=0.0927$ with $N_{it}=1/P=10.78$, for $I=8$. The same numerical experiment as in Example 1 is performed for $I=16$, and we get $P=0.0071$ with $N_{it}=1/P=140.95$, while the statistics for this case produced $N_{it}=139.78$. The same holds for $I=24$. The complete results of the experiment, with the modified impulsive noise, are given in Table \ref{TableII}.

\begin{table}[htbp]
	\centering
		\begin{tabular}{|l|r|r|r|r|r|r|}
			\hline 
			& $N_{it}$ & $SNR_{in} $ & $SNR_{in0}$ & $SNR_{out0}$ & $SNR_{out}$ &  $D$ \\ \hline
			
			$I=8$ & 10.74 &  -22.45 &  13.01 &  21.02 &  27.20 & 119.96 \\ \hline
			
			$I=16$ & 139.78 & -26.75 &   13.00 &  21.06 &   27.05 &  111.94 \\ \hline
			
			$I=24$  & 2335.67 & -27.21 &  13.01 &   20.71 &   26.31 &  103.56 \\ \hline
		\end{tabular}
	\smallskip
	\caption{Results for the case with $N=128$, $M=32$, $K=5$, with highly impulsive noise so that all its values are  outliers.}
	\label{TableII}
\end{table}

In this case (of the modified impulsive noise), we can also check the result for the SNR ratio in the final RANSAC mid-result, when the consensus is detected on a small subset with $M$ samples. Then,  $$SNR_{out0}-SNR_{in0}=10\log\bigg(\frac{M}{K}\bigg)=10\log\bigg(\frac{32}{5}\bigg)=8.06 \ \mathrm{ dB}$$
for all three considered cases, $I=8$, $I=16$, and $I=24$. This result is in complete agreement with the statistical results for these SNR values in Table \ref{TableII}. 

\smallskip
\noindent\textbf{Example 3:}  
The experiment form Example 1 is repeated with some other numbers of the available samples, $N$, sparsites, $K$, the number of impulsive disturbances, $I$, and the  samples used in the RANSAC-based CS reconstruction. The results are given in Tables \ref{TableIII} and \ref{TableIV} and further prove the efficiency of the proposed method and accuracy of the proposed SNR descriptors. 

\begin{table}[htbp]
		\centering
		\begin{tabular}{|l|r|r|r|r|r|r|}
			\hline 
			& $N_{it}$ & $SNR_{in} $ & $SNR_{in0}$ & $SNR_{out0}$ & $SNR_{out}$ &  $D$ \\ \hline
			
			$I=8$ & 5.40 &  -5.24 &  16.79  & 14.90  & 26.30 & 107.52 \\ \hline 
			
			$I=12$ &
			15.06 & -10.13  & 16.76 &  13.33 &  25.79  & 104.85 \\  \hline
			
			$I=16$  &
			99.47 & -11.42 &  16.80  & 13.10  & 25.41 &   101.41
			\\ \hline
		\end{tabular}
	\smallskip
	\caption{Results for the case with $N=128$, $M=64$, $K=12$.}
	\label{TableIII}
\end{table}

\begin{table}[htbp]
	\centering
		\begin{tabular}{|l|r|r|r|r|r|r|}
			\hline 
			& $N_{it}$ & $SNR_{in} $ & $SNR_{in0}$ & $SNR_{out0}$ & $SNR_{out}$ &  $D$ \\ \hline
			
			$I=16$ &
			5.36  & -8.63  & 16.83  & 15.79 &  29.14 & 231.65 \\ \hline
			
			$I=32$ &
			61.84 & -14.55  & 16.84  &  14.15  & 28.30 & 214.07 \\ \hline
			$I=40$  &
			261.30 & -16.90 &  16.81  &  14.42  & 28.25 & 208.48 
			\\ \hline
		\end{tabular}
	\smallskip
	\caption{Results for the case with $N=256$, $M=64$, $K=12$.}
	\label{TableIV}
\end{table}

\smallskip
\noindent\textbf{Example 4:}  
The experiment from Example 1 is repeated with the impulsive noise only, that is, when $y(n)=x(n)+\nu(n)$. As expected, for all considered numbers of outliers, the obtained results are within the computer precision accuracy.  They are given in Table \ref{TableV}. In this case, the value of $d$ should very small. We  used $d=10^{-6}$, for this experiment.

\begin{table}[htbp]
	\centering
	\begin{footnotesize}
		\begin{tabular}{|l|r|r|r|r|r|r|}
			\hline 
			& $N_{it}$ & $SNR_{in} $ & $SNR_{in0}$ & $SNR_{out0}$ & $SNR_{out}$ &  $D$ \\ \hline
			
			$I=8$ & 10.68 &  -22.44 &  316.64 &  276.25 &  282.12 & 120 \\ \hline
			
			$I=16$ & 138.21 & -26.75 &  316.67 &  271.86 &   277.52 &  112 \\ \hline
			
			$I=24$  & 2402.23 & -27.47 &  316.67 &   270.51 &   275.40 &  104 \\ \hline
		\end{tabular}
	\end{footnotesize}
	\smallskip
	\caption{Results for the case with $N=128$, $M=32$, $K=5$, with highly impulsive noise in outliers and without noise in inliers.}
	\label{TableV}
\end{table}

\section{Conclusion}
Inspired by recent advances in compressive sensing and sparse signal processing, we have developed a RANSAC-based methodology for the detection of disturbances. Upon detecting disturbance-free samples, a compressive sensing algorithm is used for the recovery of the disturbed samples, which are considered as unavailable. The presented methodology is general -- no specific assumptions have been made regarding the range of values or statistical behavior of the disturbance. The presented approach exploits the fact that disturbances degrade signal sparsity. It has been only assumed that the signals of interest exhibit sparsity in a known transformation domain. The theory has been verified on numerical examples.

\section*{Acknowledgment}
This work was supported by the Croatian Science Foundation under the project IP-2020-02-4358.

\end{document}